\documentclass{article}
\usepackage{spconf,amsmath,graphicx}
\usepackage{multirow,array,enumitem,adjustbox}
\usepackage{booktabs}
\usepackage{xcolor}

\usepackage{amssymb}
\usepackage{pifont}
\newcommand{\cmark}{\ding{51}}%
\newcommand{\xmark}{\ding{55}}%

\usepackage{hyperref}
\usepackage{cleveref}

\usepackage{etoolbox}
\patchcmd{\thebibliography}
  {\settowidth}
  {\setlength{\itemsep}{-1.5pt plus -1.5pt}\settowidth}
  {}{}
\apptocmd{\thebibliography}
  {\small}
  {}{}

\title{Large-scale Self-Supervised Speech Representation Learning  for\\Automatic Speaker Verification}
\name{\begin{tabular}{c}Zhengyang Chen$^{1,2,*}$ \thanks{$^*$Work done during an internship at Microsoft.}, Sanyuan Chen$^2$,  Yu Wu$^2$, Yao Qian$^2$,  Chengyi Wang$^2$ \\ Shujie Liu$^2$, Yanmin Qian$^1$, Michael Zeng$^2$ \end{tabular} }

\address{$^1$ MoE Key Lab of Artificial Intelligence, AI Institute
, \\
X-LANCE Lab, Department of Computer Science and Engineering, Shanghai Jiao Tong University\\
$^2$Microsoft Corporation}

\begin{document}
\ninept
\maketitle
%






\begin{abstract}
The speech representations learned from large-scale unlabeled data have shown better generalizability than those from supervised learning and thus attract a lot of interest to be applied for various downstream tasks. In this paper, we explore the limits of speech representations learned by different self-supervised objectives and datasets for automatic speaker verification (ASV), especially with a well-recognized SOTA ASV model, ECAPA-TDNN \cite{desplanques2020ecapa}, as a downstream model. The representations from all hidden layers of the pre-trained model are firstly averaged with learnable weights and then fed into the ECAPA-TDNN as input features. The experimental results on Voxceleb dataset show that the weighted average representation is significantly superior to FBank, a conventional handcrafted feature for ASV. Our best single system achieves 0.537\%, 0.569\%, and 1.180\% equal  error  rate (EER) on the three official trials of VoxCeleb1, separately. Accordingly, the ensemble system with three pre-trained models can further improve the EER to 0.479\%, 0.536\% and 1.023\%. Among the three evaluation trials, our best system outperforms the winner system \cite{zhao2021speakin} of the VoxCeleb Speaker Recognition Challenge 2021 (VoxSRC2021) on the VoxCeleb1-E trial.

\end{abstract}

\vspace{-5pt}
\section{Introduction}
\label{sec:intro}
\vspace{-5pt}

Recent years have witnessed significant improvements in automatic speaker verification (ASV) tasks. Researchers have developed various neural network architectures \cite{desplanques2020ecapa,liu2015deep, snyder2018x-vector, zeinali2019but}, training objectives \cite{xiang2019margin, zhang2017end, huang2018angular, wan2018generalized}, pooling functions \cite{okabe2018attentive, zhu2018self} to push the limits of the system performance. However, these techniques always require large-amount well-labeled data. It is a challenge to collect large-scale labeled data for real applications due to the privacy issue of speaker information. Over the past years, pre-trained models have become the de-facto standard for state-of-the-art performance on many natural language processing (NLP) tasks. Inspired by the great success of BERT \cite{devlin2018bert} and GPT \cite{radford2018improving},  a series of work in the speech community, e.g. wav2vec 2.0 \cite{baevski2020wav2vec} and HuBERT \cite{hsu2021hubert}, have been proposed to leverage large-scale unlabeled data, showing the impressive results on the automatic speech recognition (ASR) tasks. 

For the speaker verification field, many researchers have designed specific losses to train the extractor of speaker embeddings from the unlabeled data under an assumption that there is only one speaker in one utterance \cite{zhang2021contrastive, xia2021self, cai2021iterative}. Such an assumption may limit the application for un-supervised speaker verification training on the unlimited data from the internet. The Wav2Vec 2.0 \cite{baevski2020wav2vec} and HuBERT \cite{hsu2021hubert} rely less on such assumption.
These two pre-trained models have shown that they can capture phonetic structure information contained in speech and thus benefit ASR. It is an interesting research topic to probe the nature of the representations learned by different layers of pre-trained models \cite {jawahar-etal-2019-bert,DBLP:journals/corr/abs-2107-04734}. The effectiveness of Wav2vec 2.0 in a two-stage training process of pre-trained and fine-tuning has been demonstrated on both speaker verification and language recognition tasks in \cite{fan21_interspeech}. Besides, \cite{yang2021superb} introduces a benchmark to evaluate the performance of pre-trained models and shows the better performance of the speech representations learned from large-scale unlabeled data, by comparing with Fbank, on various downstream tasks including ASV. In order to minimize architecture changes and fine-tuning to solve all downstream tasks, the works above only use a simple downstream model and train the system on a small speaker verification dataset Voxceleb1 \cite{nagrani2017voxceleb} for ASV task. However, whether the speech representations can also benefit the state-of-the-art (SOTA) ASV systems is still an open question.   



In this paper, the speech representations learned from large-scale unlabeled data are extensively investigated on a benchmark dataset for speaker verification. The major contribution of this paper is four-fold as follows:
\begin{enumerate}
   \item To the best of our knowledge, it is the first attempt to use the speech representation learned from large-scale unlabeled data to improve the performance of the SOTA speaker verification model (i.e., ECAPA-TDNN \cite{desplanques2020ecapa} ) on Voxceleb dataset.
   \item Instead of using the representations only from the final layer of the pre-trained model, we employ a weighted average of the representations from all hidden layers to fully leverage the speaker-related information embedded in the whole model.
   \item We conduct a comprehensive study on the performance of pre-trained models with different learning methods, model sizes and large-scale training datasets. 
   \item A detailed analysis based on learnable weights is performed for probing layer-wise speaker information embedded in the pre-trained models.
\end{enumerate}

\vspace{-5pt}
\section{Related Work}
\label{sec:relate_work}
\vspace{-5pt}

Speech signals contain all kinds of information, such as phonetic structure, emotion, speaker indentity, etc. 
The Fbank and MFCC are the most commonly used handcrafted acoustic features, which demonstrate sound characteristics in the frequency domain. In addition, researchers have been doing lots of feature engineering to improve their performance, e.g., delta features to capture temporal dynamics of Fbank or MFCC. The authors in \cite{todisco2016articulation} combined the articulation rate filter with the constant Q cepstral coefficients (CQCCs) \cite{todisco2017constant} in the speaker verification task and achieved significant improvement compared to MFCC baseline. In order to make better use of the powerful learning ability of neural networks, Mirco et al. \cite{sincnet} and Jee-weon et al. \cite{jung2019rawnet} have tried to used convolution neural network to learn task-specific features from raw audio signals and achieved comparable performance with handcrafted feature.

Recently, speech representation learning by leveraging unlabeled data is gradually emerging. It is commonly believed that the pre-trained models by self-supervised learning have a good generalizability and a simple classifier added on the top of the representations from these pre-trained models can obtain decent performance for many downstream tasks, even with a limited amount of labeled data. Self-supervised learning for speech representations can be categorized into three approaches: 1) reconstruction learning aims to reconstruct the original input using information extracted from past time steps or masked inputs; 2) Contrastive learning learns high-level representations by solving a contrastive task in the latent embedding space; 3) multi-task learning with multiple objectives and multiple inputs. A review of these approaches is given in \cite{yang2021superb}.


\begin{figure}[ht]
\centering
\includegraphics[width=0.47\textwidth]{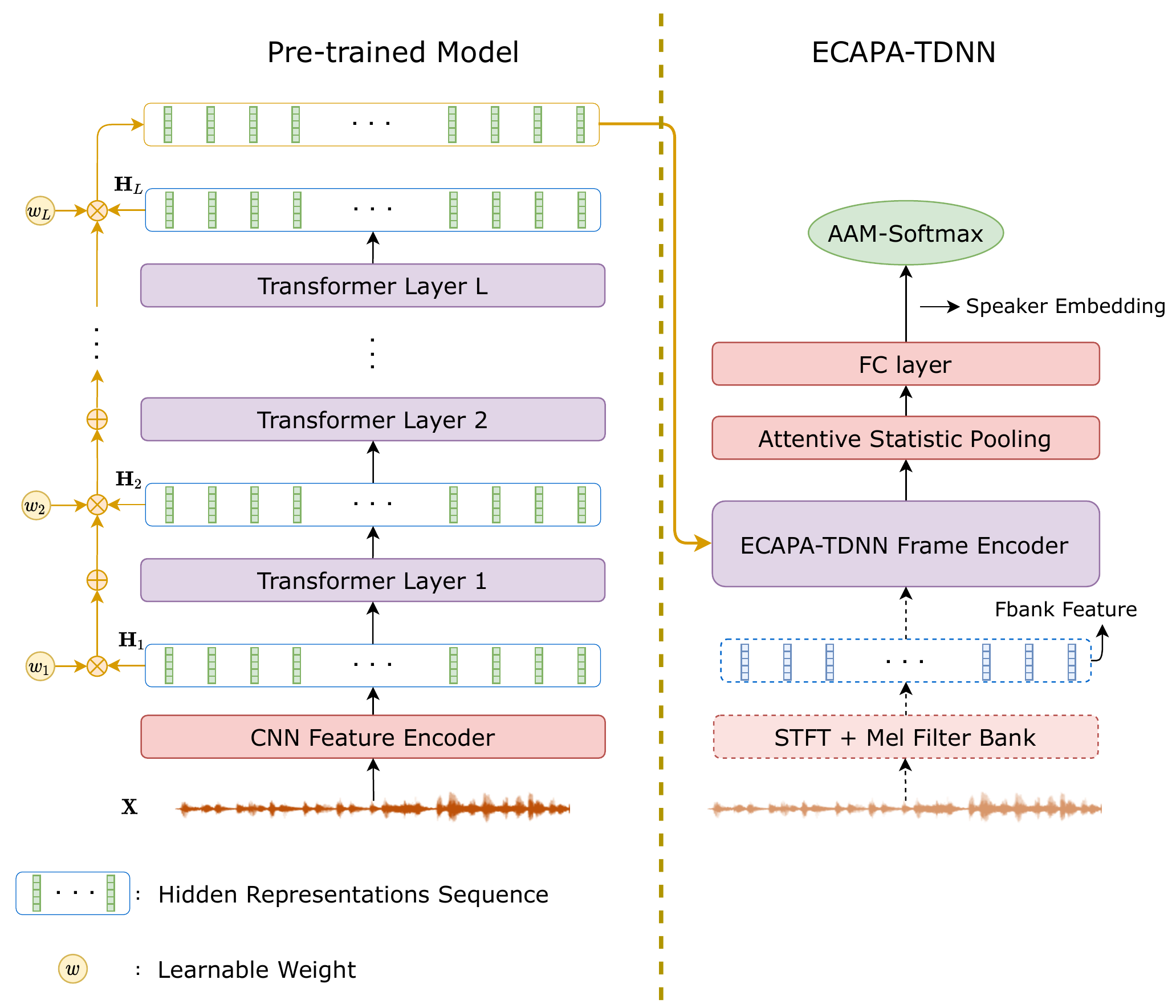}
\caption{\textbf{Leverage Representations from Pre-trained Model}}
\label{fig:model}
\vspace{-15pt}
\end{figure}

\section{Methods}
\label{sec:method}

\subsection{Pre-train for Representation Learning}
In this study, we leverage the representations from Wav2Vec 2.0 \cite{baevski2020wav2vec}, HuBERT \cite{hsu2021hubert} and UniSpeech-SAT \cite{chen2021unispeech} to do speaker verification task. These three models use different methods to learn the feature representation. The Wav2Vec 2.0 model uses a contrastive loss to distinguish a true speech segment from negatives. 
The goal of HuBERT is to predict the weak supervised label for the mask frames.
UniSpeech-SAT integrates an utterance-wise contrastive loss into Hubert-like representation learning that forces speaker-related information into the learned representation. 
Despite the different training objectives for the pre-trained models described above, they share the similar model structures. As shown in the left part of Figure \ref{fig:model}, these three pre-trained models all consist of a convolutional feature extractor and a deep transformer \cite{vaswani2017attention} network as the encoder. Mathematically, given an input wavform  $\mathbf{X} = \{x_1 \ldots x_N \}_{t=1}^N$ where $N$ is the number of sampling points, the CNN feature encoder convolves the sample points to a sequence of feature vector, $\mathbf{H}_0 = \{\textbf{h}_{1,0} \ldots \textbf{h}_{T,0} \}_{t=1}^T$. Then the sequence of feature vector is fed to the Transformer model, yielding a hidden state for each frame at the $l$-th layer $\mathbf{H}_l = \{\textbf{h}_{1,l} \ldots \textbf{h}_{T,l} \}_{t=1}^T$, where $l \in \{1, \ldots ,L\}$.

\vspace{-10pt}
\subsection{Leverage Representations from pre-trained Model}

\subsubsection{Downstream Speaker Verification Model}
\vspace{-5pt}

In \cite{fan21_interspeech}, the authors added an average pooling layer and a fully connected layer with a task-specific loss on the top of pre-trained models and achieved comparable results with the systems using handcrafted features. In \cite{yang2021superb}, x-vector \cite{snyder2018x-vector} is used as the downstream model. To push the limit of the performance of the downstream task, we use the state-of-the-art speaker verification system ECAPA-TDNN \cite{desplanques2020ecapa} as the downstream model. Compared to x-vector, ECAPA-TDNN has a more advanced design, e.g. Squeeze-Excitation Res2Blocks \cite{hu2018squeeze, gao2019res2net} and multi-layer feature aggregation, which significantly improves system performance.  The brief structural framework of ECAPA-TDNN is shown as the right part of Figure \ref{fig:model}. The model takes the sequence of the Fbank feature as input. Then, the frame encoder extracts speaker information from each input frame and the statistic pooling layer transforms the variable length input sequence to fix-dimensional representation. Finally, a fully connected (FC) layer is added to extract speaker embedding.
To leverage the representations learned from the pre-trained models, we can replace Fbank with the last-layer outputs of pre-trained models and feed them into the ECAPA-TDNN. 

\begin{table*}[th!]
\centering
\caption{\textbf{The detailed information of pre-trained models used in our experiments and down-stream task model.} For the UniSpeech-SAT\_* models, we use Librivox (60k hrs), VoxPopuli (24k hrs) and Gigaspeech (10k hrs, English) to form the 90k hours training data. The layer \# column only counts the transformer layer in the pre-training model.}
\begin{adjustbox}{width=.98\textwidth,center}
\begin{tabular}{cccccc}
    \toprule 
  
    \multirow{2}{*}{Pre-training/Down-stream Model} & \multirow{2}{*}{Layer \#} & \multirow{2}{*}{Parameter \#} & \multicolumn{3}{c}{Training Data}  \\
    \cline{4-6}
    & & & Duration & Sources & Language \\
    \hline
    HuBERT\_Base         & 12 & $\sim$95M & 960 hrs & Librispeech & English \\
    HuBERT\_Large        & 24 & $\sim$316M & 60k hrs & Librivox & English \\
    Wav2Vec2.0\_Large (XLSR)       & 24 & $\sim$316M & 56k hrs & Multilingual LibriSpeech, CommonVoice, BABLE & Over 36 languages \\
    UniSpeech-SAT\_Base  & 12 & $\sim$95M & 94k hrs & Librivox, VoxPopuli, Gigaspeech & English \\
    UniSpeech-SAT\_Large & 24 & $\sim$316M & 94k hrs & Librivox, VoxPopuli, Gigaspeech & English \\
    \hline
    ECAPA-TDNN (small) \cite{desplanques2020ecapa}   & - & $\sim$6M & 2.36k hrs & Voxceleb2 (Youtube) & Multi-Lingual (mostly English) \\
    \hline
   
    \bottomrule
\end{tabular}
\label{table:pretrain_model}
\end{adjustbox}
\vspace{-10pt}
\end{table*}

\vspace{-10pt}
\subsubsection{Explore Speaker Information in Pre-trained Model}
\label{sssec:method_speaker_info}
\vspace{-5pt}

The pre-trained model, which has seen tons of audio data, should have good generalization for various downstream tasks. However, the results in \cite{fan21_interspeech} didn't show the superiority of the pre-trained representation compared to handcrafted feature. The objectives of the most pre-trained tasks are not directly related to speaker recognition. The layers close to the final objectives will contain more information related to the training loss. It could be better to discover the speaker information from the low layers of the pre-trained model.

Here, similar to the implementation in \cite{yang2021superb, pepino21_interspeech}, we introduce a learnable weight, $w_l$, for hidden states $\mathbf{H}_l, l \in \{0, \ldots ,L\}$ from each layer in pre-trained model. Rather feeding the outputs from the last layer of the pre-trained model, i.e. $\mathbf{H}_L$, to the downstream model, we weighted average the hidden states of each layer to generate the frame representation $\textbf{o}_t = \sum_{l=0}^{L} w_l \cdot h_{l,t}$. Then, we replace the Fbank feature fed into the ECAPA-TDNN with the weighted average representations to extract speaker embedding $\textbf{e}$:

\vspace{-5pt}
\begin{equation}
\textbf{e} = \text{ECAPA-TDNN}(\textbf{o}_1 \ldots \textbf{o}_T)
\end{equation}
\vspace{-10pt}

Same as the implementation in \cite{desplanques2020ecapa}, we also use the additive angular margin (AAM) \cite{deng2019arcface} loss in the training process for model optimization.

The training pipeline is mainly divided into two stages. In the first stage, the pre-trained model is fixed. We only update the ECAPA-TDNN and the weight $w$ for all the hidden states. Then, we fine-tune all the parameters for pre-trained model and ECAPA-TDNN.

\vspace{-5pt}
\section{Experimental Setup}
\label{sec:exp_setup}
\vspace{-5pt}

To analyze the effectiveness of pre-trained model representation for speaker verification task, we trained and evaluated the downstream speaker verification model using Voxceleb1 \cite{nagrani2017voxceleb} and Voxceleb2 \cite{chung2018voxceleb2} datasets.  All three official trial lists Vox1-O, Vox1-E and Vox1-H are used to evaluate the system performance. When implementing our baseline models using the handcrafted acoustic feature, we extract 40-dimensional Fbank feature with 25ms window size and 10ms frame shift. We didn't do voice activity detection (VAD) processing for the Voxceleb data. Besides, we also did data augmentation for the training data using the MUSAN \cite{musan2015} noise and RIR \footnote{\url{https://www.openslr.org/28/}} reverberation with probability 0.6 in online mode.

The detailed information about the pre-trained models used in our experiments and the speaker verification downstream models is listed in Table \ref{table:pretrain_model}. The HuBERT\_Base, HuBERT\_Large and Wav2vec2.0\_Large (XLSR) models are released by Fairseq sequence modeling toolkit \footnote{\url{https://github.com/pytorch/fairseq}}. The results in \cite{yang2021superb} show that the Wav2vec2.0\_Base performed worse than HuBERT\_Base on speaker-related task and we didn't use it here. UniSpeech-SAT is a model proposed recently, which explicitly models the speaker information in pre-training process. It introduces utterance contrastive loss to model the single speaker information, where the positive instances are hidden states in the same utterance while the negative instances are hidden states in other utterances. Moreover, UniSpeech-SAT uses more synthesis or public available data compared to HuBERT. For downsteam task model, we use the small ECAPA-TDNN in \cite{desplanques2020ecapa}.


We trained all the models with Additive Angular Margin Loss (AAM) \cite{deng2019arcface} and set the margin to 0.2. During the training process, we randomly sampled 3s segment from each utterance to construct training batch. For the two-stage training pipeline described in section \ref{sssec:method_speaker_info}, we first fixed the pre-trained model and trained for 10 epochs. Then, we fine-tuned all the parameters for another 5 epochs. Besides, to further improve our best system, we did large margin fine-tuning \cite{thienpondt2021idlab} by randomly sampling 6s segments and set the AAM margin to 0.5 to train extra 2 epochs.


During the evaluation, we use the cosine score to measure the similarity for trial pairs. We also use the adaptive s-norm \cite{karam2011towards,cumani2011comparison} to normalize the scores in our experiment. The embeddings extracted from the training set are averaged according to the speaker label and used as the imposter cohort. We set the imposter cohort size to 600 in our experiment. When doing quality-aware score calibration \cite{thienpondt2021idlab}, we randomly generated 30k trials based on the voxceleb2 test set to train our calibration model.

\begin{table}[ht!]
\footnotesize
\centering
\caption{\textbf{Comparison with traditional acoustic feature based on Voxceleb1.} Here, we trained all the models on Voxceleb1 dev set and evaluated on Vox1-O trial. We fixed pre-trained model in the training process and only use them to extract speech representation.}
\begin{adjustbox}{width=.45\textwidth,center}
\begin{tabular}{cccc}
    \toprule 
  
     Feature & Aug & Pretrain Feature & Vox1-O EER (\%) \\
    
    \hline
    Fbank & \xmark & - & 3.899 \\
    HuBERT\_Base & \xmark  & Last & 3.691 \\
    HuBERT\_Base &  \xmark & Hidden & 2.117 \\
    \midrule
    Fbank & \cmark & -  & 2.371 \\
    HuBERT\_Base & \cmark & Last & 3.079 \\
    HuBERT\_Base & \cmark & Hidden & 1.861 \\
    UniSpeech-SAT\_Base  & \cmark  & Hidden  & 1.632 \\
    
    \bottomrule
\end{tabular}
\label{table:res_baseline}
\end{adjustbox}
\vspace{-15pt}
\end{table}

  

\label{sec:res_analysis}

\section{Evaluation Results}
\label{ssec:fbank_compare}

\vspace{-5pt}

\subsection{Comparison with Handcrafted Acoustic Feature}

First, we will compare the speech representations extracted from pre-trained models with the commonly used handcrafted feature. The experiments in \cite{fan21_interspeech} have shown that Wav2Vec 2.0 pre-trained models contain speaker information and can achieve comparable performance with the handcrafted acoustic feature. Different from \cite{fan21_interspeech}, in our experiments, we directly replaced the handcrafted feature fed to the speaker verification model ECAPA-TDNN with the representations from pre-trained models. Besides, we explored to leverage the representations from pre-trained models in two different ways, using the representation from the last layer or weighted averaging all the hidden representations. The results are shown in Table \ref{table:res_baseline}. From the upper part of the table, we find that the last layer representation and all hidden layers'    representation from the pre-trained model both perform better than the handcrafted feature Fbank. Encouragingly, the performance of weighted averaging hidden representation exceeds the Fbank by a very large margin (46\% relatively). Then, we augment the training data and the results are listed in the bottom part of Table \ref{table:res_baseline}. With data augmentation, all the results are further improved and the weighted averaging hidden representations also shows superiority over the Fbank feature. For the experiments in the following sections, we will use the weighted average hidden representations for pre-trained model and augment the training data.  

\begin{table*}[th!]
\footnotesize
\centering
\caption{\textbf{Results with different pre-trained models and different training strategies.} Here, we did data augmentation for all the experiments. In the last Ensemble line, we weighted average the scores of our best three systems after score calibration. The weight in the score average is decided according to the performance of the single system. Besides, the Large model performs much better than the Base model and we only list a part of the results for the Base model because of the space limit.
}
\begin{adjustbox}{width=.75\textwidth,center}
\begin{tabular}{c|c|c|c|c|ccc} 
    \toprule 
  
    \multirow{2}{*}{Train Data} & Large Margin & Score & \multirow{2}{*}{Fix Pretrain} &  \multirow{2}{*}{Feature}  & \multicolumn{3}{c}{EER (\%)} \\
    \cline{6-8}
    & Finetune & Calibration & & & Vox1-O & Vox1-E & Vox1-H \\
    \hline
    \multirow{9}{*}{Vox1\_dev} & \multirow{9}{*}{\xmark} & \multirow{9}{*}{\xmark} & - & Fbank & 2.371 & - & - \\
    &  &  & \cmark & UniSpeech-SAT\_Base  & 1.632 & - & - \\
    &  &  & \cmark & HuBERT\_Large & 1.436 & - & - \\
    &  &  & \cmark & Wav2Vec2.0\_Large (XLSR) & 1.362 & - & - \\
    &  &  & \cmark & UniSpeech-SAT\_Large & 1.249 & - & - \\
    \cline{4-8}
    &  &  & \xmark & UniSpeech-SAT\_Base  & 1.611 & - & - \\
    &  &  & \xmark & HuBERT\_Large & 1.404 & - & - \\
    &  &  & \xmark & Wav2Vec2.0\_Large (XLSR) & 1.335 & - & - \\
    &  &  & \xmark & UniSpeech-SAT\_Large & 1.218 & - & - \\
    \hline
    \multirow{18}{*}{Vox2\_dev} & \xmark & \xmark & - & Fbank  (ECAPA-TDNN small \cite{desplanques2020ecapa})             & 1.010 & 1.240 & 2.320 \\
    & \xmark & \xmark & - & Fbank  (ECAPA-TDNN large \cite{desplanques2020ecapa})                & 0.870 &  1.120 & 2.120 \\
    & \xmark & \xmark & - & Fbank                & 1.080 & 1.200 & 2.127 \\
    & \xmark & \xmark & \cmark & UniSpeech-SAT\_Base  & 1.095 & 1.152 & 2.221 \\
    & \xmark & \xmark & \cmark & HuBERT\_Large & 0.914 & 0.948 & 1.759 \\
    & \xmark & \xmark & \cmark & Wav2Vec2.0\_Large (XLSR)  & 1.021 & 0.962 & 1.782 \\
    & \xmark & \xmark & \cmark & UniSpeech-SAT\_Large  & 0.750 & 0.813 & 1.649 \\
    \cline{2-8}
    & \xmark & \xmark & \xmark & UniSpeech-SAT\_Base  & 1.005 & 0.933 & 1.866 \\
    & \xmark & \xmark & \xmark & HuBERT\_Large & 0.814 & 0.777 & 1.505 \\
    & \xmark & \xmark & \xmark & Wav2Vec2.0\_Large (XLSR)  & 0.803 & 0.729 & 1.394 \\
    & \xmark & \xmark & \xmark & UniSpeech-SAT\_Large  & 0.696 & 0.685 & 1.433 \\
    \cline{2-8}
    & \cmark & \xmark & \xmark & HuBERT\_Large & 0.723 & 0.706 & 1.317 \\
    & \cmark & \xmark & \xmark & Wav2Vec2.0\_Large (XLSR)  & 0.734 & 0.677 & 1.235 \\
    & \cmark & \xmark & \xmark & UniSpeech-SAT\_Large  & 0.633 & 0.625 & 1.294 \\
    \cline{2-8}
    & \cmark & \cmark & \xmark & HuBERT\_Large & 0.590 & 0.654 & 1.227 \\
    & \cmark & \cmark & \xmark & Wav2Vec2.0\_Large (XLSR)  & 0.585 & 0.625 & \textbf{1.138} \\
    & \cmark & \cmark & \xmark & UniSpeech-SAT\_Large  & \textbf{0.537} & \textbf{0.569} & 1.180 \\
    \cline{2-8}
    & \cmark & \cmark & \xmark & Ensemble  & 0.479 & 0.536 & 1.023 \\
    \bottomrule
\end{tabular}
\label{tab:vox2_res}
\end{adjustbox}
\end{table*}

\vspace{-5pt}
\subsection{Comparison among Different Pre-trained Models}

To further improve the effectiveness of the representations from pre-trained models, we trained the model on a larger dataset, Voxceleb2\_dev, and compared different pre-trained models and training strategies. All the results are shown in Table \ref{tab:vox2_res}. The results show that all the large models perform better than Fbank feature on both Vox1\_dev and Vox2\_dev setup. When we unfix the pre-trained model and jointly fine-tune the pre-trained model and downstream model, further improvements can be achieved. The improvement from pre-trained model fine-tuning is more obvious on Vox2\_dev setup than Vox1\_dev setup. Besides, the Wav2vec2.0\_Large (XLSR) and UniSpeech-SAT\_Large pre-trained models perform better than the HuBERT\_Large after fine-tuning. As shown in table \ref{table:pretrain_model}, the training set size of the Wav2vec2.0\_Large (XLSR) and HuBERT\_Large is comparable. However, the training data for Wav2vec2.0\_Large (XLSR) is more diverse and more matched with Voxceleb data, enabling it to be more suitable for this downstream task. Moreover, the UniSpeech-SAT\_Large model with more training data performs the best among most of the trials. Compared to Fbank feature, representations from this model achieved $\sim$30\% relative EER improvement on all three trials for the Voxceleb1 evaluation set.

In \cite{thienpondt2021idlab}, the authors introduced a large margin fine-tuning strategy and quality-aware score calibration to the speaker verification task and achieved impressive improvement. Here, we also leverage these two strategies in our experiments to push the performance limit. The corresponding results are listed at the bottom part in Table \ref{tab:vox2_res}. With these two strategies, our best system exceeds the state-of-the-art system \cite{zhao2021speakin} (Vox1-O: 0.461, Vox1-E: 0.634, Vox1-H: 0.993) in VoxSRC challenge 2021 on Vox-E trial.

\begin{figure}[ht]
\centering
\includegraphics[width=0.45\textwidth]{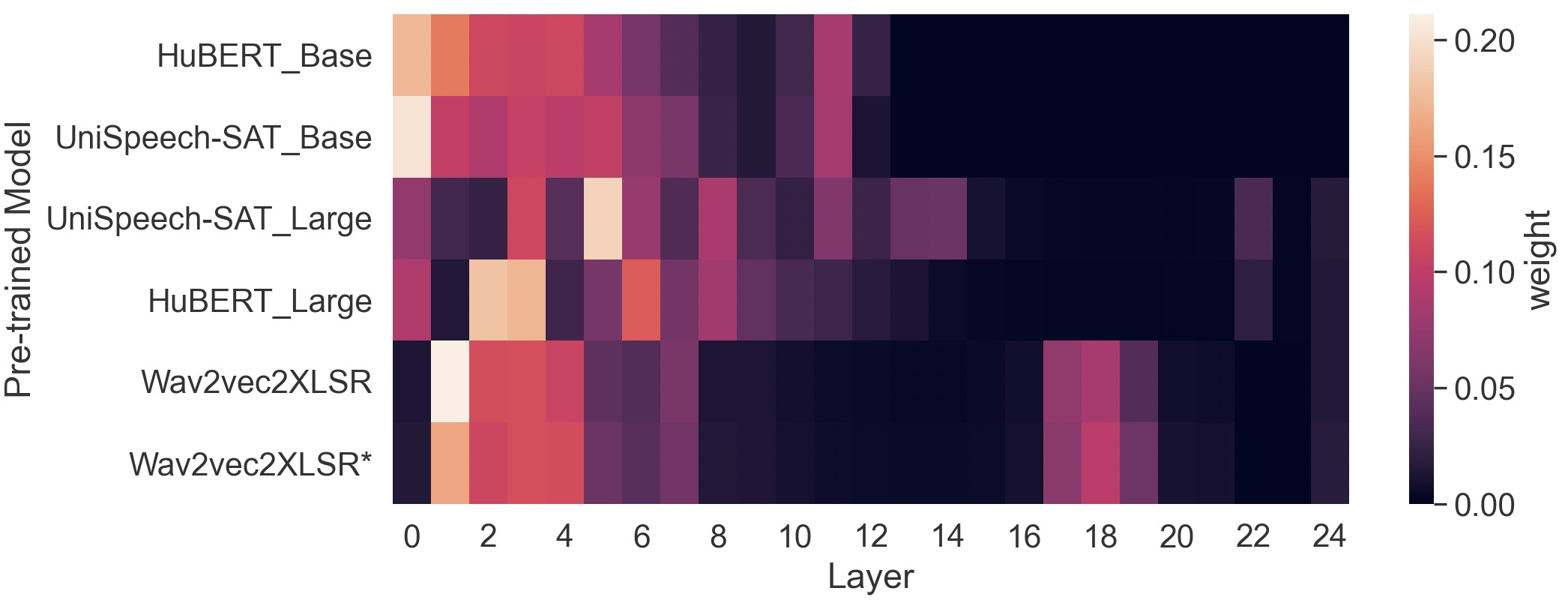}
\vspace{-10pt}
\caption{\textbf{The visualization of the normalized weight values in the proposed architecture show as Figure \ref{fig:model}}. The output from the layer 0 corresponds to the transformer input. * in the figure means that the pre-trained model is unfixed during downsteam task training. It should be noted that base models only have 12 layers and other models have 24 layers.}
\label{fig:weight_visualization}
\end{figure}

\subsection{Analysis Speaker Information in Pre-trained Model}

The results in Section \ref{ssec:fbank_compare} have shown that it is better to leverage the representations from all the hidden layers rather than the last layer. Thus, it could be necessary and meaningful to explore which layer contains more speaker information than the others. We visualize the normalized weight value for all the layers' output in Figure \ref{fig:weight_visualization}. The figure shows that the speaker information at the lower layers of pre-trained models is more discriminative than those at the higher layers for ASV task . This phenomenon is reasonable because the training objectives for the pre-trained models used in our experiments are more related to the speech recognition task. For large pre-trained models in our experiments, i.e. UniSpeech-SAT\_Large, HuBERT\_Large and Wav2vec2.0\_Large (XLSR), the learned weights assigned to the higher layers are much smaller than those of lower layers, which indicates that we might be able to directly throw away these higher layers to reduce model size.

\vspace{-5pt}
\section{Conclusion}
\vspace{-5pt}
In this paper, we leverage the representations extracted from pre-trained models trained on large-scale unlabeled data in speaker verification task.  In our experiments, we first compared such representations with handcrafted Fbank feature and verify the superiority of pre-trained representations. To comprehensively explore speaker information in the pre-trained model, we make the model learn the weights automatically for all the hidden states of the pre-trained model and achieve significant performance improvement compared to the baseline. By visualizing the learned weights, we find the lower layers of the pre-trained model can capture more speaker-related information than those of higher layers. Despite the significant improvement benefiting from the pre-trained model, there is still a relatively small performance gap (on two evaluation sets) between our system and the best system \cite{zhao2021speakin} in the VoxSRC2021 challenge, which has a more aggressive augmentation strategy and dedicated training objectives. In the future, we will incorporate the better training setup in \cite{zhao2021speakin} for our system to further push the limit of speaker verification performance. 




\bibliographystyle{IEEEtran}
{\small \bibliography{refs}}

\begin{thebibliography}{10}
\providecommand{\url}[1]{#1}
\csname url@samestyle\endcsname
\providecommand{\newblock}{\relax}
\providecommand{\bibinfo}[2]{#2}
\providecommand{\BIBentrySTDinterwordspacing}{\spaceskip=0pt\relax}
\providecommand{\BIBentryALTinterwordstretchfactor}{4}
\providecommand{\BIBentryALTinterwordspacing}{\spaceskip=\fontdimen2\font plus
\BIBentryALTinterwordstretchfactor\fontdimen3\font minus
  \fontdimen4\font\relax}
\providecommand{\BIBforeignlanguage}[2]{{%
\expandafter\ifx\csname l@#1\endcsname\relax
\typeout{** WARNING: IEEEtran.bst: No hyphenation pattern has been}%
\typeout{** loaded for the language `#1'. Using the pattern for}%
\typeout{** the default language instead.}%
\else
\language=\csname l@#1\endcsname
\fi
#2}}
\providecommand{\BIBdecl}{\relax}
\BIBdecl

\bibitem{desplanques2020ecapa}
B.~Desplanques, J.~Thienpondt, and K.~Demuynck, ``{ECAPA-TDNN: Emphasized
  Channel Attention, Propagation and Aggregation in TDNN Based Speaker
  Verification},'' in \emph{Proc. Interspeech 2020}, 2020, pp. 3830--3834.

\bibitem{zhao2021speakin}
M.~Zhao, Y.~Ma, M.~Liu, and M.~Xu, ``The speakin system for voxceleb speaker
  recognition challange 2021,'' \emph{arXiv preprint arXiv:2109.01989}, 2021.

\bibitem{liu2015deep}
Y.~Liu, Y.~Qian, N.~Chen, T.~Fu, Y.~Zhang, and K.~Yu, ``Deep feature for
  text-dependent speaker verification,'' \emph{Speech Communication}, vol.~73,
  pp. 1--13, 2015.

\bibitem{snyder2018x-vector}
D.~Snyder, D.~Garcia-Romero, G.~Sell, D.~Povey, and S.~Khudanpur, ``X-vectors:
  Robust dnn embeddings for speaker recognition,'' in \emph{Proc. IEEE ICASSP
  2018}.\hskip 1em plus 0.5em minus 0.4em\relax IEEE, 2018, pp. 5329--5333.

\bibitem{zeinali2019but}
H.~Zeinali, S.~Wang, A.~Silnova, P.~Mat{\v{e}}jka, and O.~Plchot, ``But system
  description to voxceleb speaker recognition challenge 2019,'' \emph{arXiv
  preprint arXiv:1910.12592}, 2019.

\bibitem{xiang2019margin}
X.~Xiang, S.~Wang, H.~Huang, Y.~Qian, and K.~Yu, ``Margin matters: Towards more
  discriminative deep neural network embeddings for speaker recognition,'' in
  \emph{Proc. APSIPA ASC 2019}.\hskip 1em plus 0.5em minus 0.4em\relax IEEE,
  2019, pp. 1652--1656.

\bibitem{zhang2017end}
C.~Zhang and K.~Koishida, ``End-to-end text-independent speaker verification
  with triplet loss on short utterances.'' in \emph{Interspeech}, 2017, pp.
  1487--1491.

\bibitem{huang2018angular}
Z.~Huang, S.~Wang, and K.~Yu, ``Angular softmax for short-duration
  text-independent speaker verification.'' in \emph{Interspeech}, 2018, pp.
  3623--3627.

\bibitem{wan2018generalized}
L.~Wan, Q.~Wang, A.~Papir, and I.~L. Moreno, ``Generalized end-to-end loss for
  speaker verification,'' in \emph{Proc. IEEE ICASSP 2018}.\hskip 1em plus
  0.5em minus 0.4em\relax IEEE, 2018, pp. 4879--4883.

\bibitem{okabe2018attentive}
K.~Okabe, T.~Koshinaka, and K.~Shinoda, ``Attentive statistics pooling for deep
  speaker embedding,'' \emph{arXiv preprint arXiv:1803.10963}, 2018.

\bibitem{zhu2018self}
Y.~Zhu, T.~Ko, D.~Snyder, B.~Mak, and D.~Povey, ``Self-attentive speaker
  embeddings for text-independent speaker verification.'' in
  \emph{Interspeech}, vol. 2018, 2018, pp. 3573--3577.

\bibitem{devlin2018bert}
J.~Devlin, M.-W. Chang, K.~Lee, and K.~Toutanova, ``Bert: Pre-training of deep
  bidirectional transformers for language understanding,'' \emph{arXiv preprint
  arXiv:1810.04805}, 2018.

\bibitem{radford2018improving}
A.~Radford, K.~Narasimhan, T.~Salimans, and I.~Sutskever, ``Improving language
  understanding by generative pre-training,'' 2018.

\bibitem{baevski2020wav2vec}
A.~Baevski, H.~Zhou, A.~Mohamed, and M.~Auli, ``wav2vec 2.0: A framework for
  self-supervised learning of speech representations,'' \emph{arXiv preprint
  arXiv:2006.11477}, 2020.

\bibitem{hsu2021hubert}
W.-N. Hsu, B.~Bolte, Y.-H.~H. Tsai, K.~Lakhotia, R.~Salakhutdinov, and
  A.~Mohamed, ``Hubert: Self-supervised speech representation learning by
  masked prediction of hidden units,'' \emph{arXiv preprint arXiv:2106.07447},
  2021.

\bibitem{zhang2021contrastive}
H.~Zhang, Y.~Zou, and H.~Wang, ``Contrastive self-supervised learning for
  text-independent speaker verification,'' in \emph{Proc. IEEE ICASSP
  2021}.\hskip 1em plus 0.5em minus 0.4em\relax IEEE, 2021, pp. 6713--6717.

\bibitem{xia2021self}
W.~Xia, C.~Zhang, C.~Weng, M.~Yu, and D.~Yu, ``Self-supervised text-independent
  speaker verification using prototypical momentum contrastive learning,'' in
  \emph{Proc. IEEE ICASSP 2021}.\hskip 1em plus 0.5em minus 0.4em\relax IEEE,
  2021, pp. 6723--6727.

\bibitem{cai2021iterative}
D.~Cai, W.~Wang, and M.~Li, ``An iterative framework for self-supervised deep
  speaker representation learning,'' in \emph{Proc. IEEE ICASSP 2021}.\hskip
  1em plus 0.5em minus 0.4em\relax IEEE, 2021, pp. 6728--6732.

\bibitem{jawahar-etal-2019-bert}
G.~Jawahar, B.~Sagot, and D.~Seddah, ``What does {BERT} learn about the
  structure of language?'' in \emph{Proc. ACL}, Jul. 2019, pp. 3651--3657.

\bibitem{DBLP:journals/corr/abs-2107-04734}
A.~Pasad, J.~Chou, and K.~Livescu, ``Layer-wise analysis of a self-supervised
  speech representation model,'' \emph{CoRR}, vol. abs/2107.04734, 2021.

\bibitem{fan21_interspeech}
Z.~Fan, M.~Li, S.~Zhou, and B.~Xu, ``{Exploring wav2vec 2.0 on Speaker
  Verification and Language Identification},'' in \emph{Proc. Interspeech
  2021}, 2021, pp. 1509--1513.

\bibitem{yang2021superb}
S.-w. Yang, P.-H. Chi, Y.-S. Chuang, C.-I.~J. Lai, K.~Lakhotia, Y.~Y. Lin,
  A.~T. Liu, J.~Shi, X.~Chang, G.-T. Lin \emph{et~al.}, ``Superb: Speech
  processing universal performance benchmark,'' \emph{arXiv preprint
  arXiv:2105.01051}, 2021.

\bibitem{nagrani2017voxceleb}
A.~Nagrani, J.~S. Chung, and A.~Zisserman, ``Voxceleb: a large-scale speaker
  identification dataset,'' \emph{arXiv preprint arXiv:1706.08612}, 2017.

\bibitem{todisco2016articulation}
M.~Todisco, H.~Delgado, and N.~W. Evans, ``Articulation rate filtering of cqcc
  features for automatic speaker verification.'' in \emph{Interspeech}, 2016,
  pp. 3628--3632.

\bibitem{todisco2017constant}
M.~Todisco, H.~Delgado, and N.~Evans, ``Constant q cepstral coefficients: A
  spoofing countermeasure for automatic speaker verification,'' \emph{Computer
  Speech \& Language}, vol.~45, pp. 516--535, 2017.

\bibitem{sincnet}
M.~Ravanelli and Y.~Bengio, ``Speaker recognition from raw waveform with
  sincnet,'' in \emph{Proc. IEEE SLT}.\hskip 1em plus 0.5em minus 0.4em\relax
  IEEE, 2018, pp. 1021--1028.

\bibitem{jung2019rawnet}
J.-w. Jung, H.-S. Heo, J.-h. Kim, H.-j. Shim, and H.-J. Yu, ``Rawnet: Advanced
  end-to-end deep neural network using raw waveforms for text-independent
  speaker verification,'' \emph{arXiv preprint arXiv:1904.08104}, 2019.

\bibitem{chen2021unispeech}
S.~Chen, Y.~Wu, C.~Wang, Z.~Chen, Z.~Chen, S.~Liu, J.~Wu, Y.~Qian, F.~Wei,
  J.~Li \emph{et~al.}, ``Unispeech-sat: Universal speech representation
  learning with speaker aware pre-training,'' \emph{arXiv preprint
  arXiv:2110.05752}, 2021.

\bibitem{vaswani2017attention}
A.~Vaswani, N.~Shazeer, N.~Parmar, J.~Uszkoreit, L.~Jones, A.~N. Gomez,
  {\L}.~Kaiser, and I.~Polosukhin, ``Attention is all you need,'' in
  \emph{Proc. NIPS}, 2017, pp. 5998--6008.

\bibitem{hu2018squeeze}
J.~Hu, L.~Shen, and G.~Sun, ``Squeeze-and-excitation networks,'' in \emph{Proc.
  CVPR}, 2018, pp. 7132--7141.

\bibitem{gao2019res2net}
S.~Gao, M.-M. Cheng, K.~Zhao, X.-Y. Zhang, M.-H. Yang, and P.~H. Torr,
  ``Res2net: A new multi-scale backbone architecture,'' \emph{IEEE transactions
  on pattern analysis and machine intelligence}, 2019.

\bibitem{pepino21_interspeech}
L.~Pepino, P.~Riera, and L.~Ferrer, ``{Emotion Recognition from Speech Using
  wav2vec 2.0 Embeddings},'' in \emph{Proc. Interspeech 2021}, 2021, pp.
  3400--3404.

\bibitem{deng2019arcface}
J.~Deng, J.~Guo, N.~Xue, and S.~Zafeiriou, ``Arcface: Additive angular margin
  loss for deep face recognition,'' in \emph{Proc. CVPR}, 2019, pp. 4690--4699.

\bibitem{chung2018voxceleb2}
J.~S. Chung, A.~Nagrani, and A.~Zisserman, ``Voxceleb2: Deep speaker
  recognition,'' \emph{arXiv preprint arXiv:1806.05622}, 2018.

\bibitem{musan2015}
D.~Snyder, G.~Chen, and D.~Povey, ``{MUSAN}: {A} {M}usic, {S}peech, and {N}oise
  {C}orpus,'' 2015, arXiv:1510.08484v1.

\bibitem{thienpondt2021idlab}
J.~Thienpondt, B.~Desplanques, and K.~Demuynck, ``The idlab voxsrc-20
  submission: Large margin fine-tuning and quality-aware score calibration in
  dnn based speaker verification,'' in \emph{Proc. IEEE ICASSP 2021}.\hskip 1em
  plus 0.5em minus 0.4em\relax IEEE, 2021, pp. 5814--5818.

\bibitem{karam2011towards}
Z.~N. Karam, W.~M. Campbell, and N.~Dehak, ``Towards reduced false-alarms using
  cohorts,'' in \emph{Proc. IEEE ICASSP 2011}.\hskip 1em plus 0.5em minus
  0.4em\relax IEEE, 2011, pp. 4512--4515.

\bibitem{cumani2011comparison}
S.~Cumani, P.~D. Batzu, D.~Colibro, C.~Vair, P.~Laface, and V.~Vasilakakis,
  ``Comparison of speaker recognition approaches for real applications.'' in
  \emph{INTERSPEECH}, 2011, pp. 2365--2368.

\end{thebibliography}

\end{document}